\documentclass[prb,twocolumn,draft,amsmath,showpacs]{revtex4}
\usepackage{graphics}
\begin{document}
%%%%%%%%%%%%%%%%%%%%%%%%%%%%%%%%%%%%%%%%%%
%\bibliographystyle{prsty}

\input epsf

\title {Influence of carbon substitution on the heat transport in 
single crystalline MgB$_2$}

\author {A.V. Sologubenko, N.D. Zhigadlo, S. M. Kazakov, J. Karpinski, H.R. Ott}
\affiliation{Laboratorium f\"ur Festk\"orperphysik, ETH H\"onggerberg,
CH-8093 Z\"urich, Switzerland}

\date{\today}

\begin{abstract}
We report data on the thermal conductivity $\kappa(T,H)$ in the basal plane of hexagonal single-crystalline and superconducting Mg(B$_{1-x}$C$_x$)$_{2}$ ($x= 0.03, 0.06$) at temperatures between 0.5 and 50 K, and in external magnetic fields $H$ between 0 and 50~kOe. The substitution of carbon for boron leads to a considerable reduction of the electronic heat transport, while the phonon thermal conductivity seems to be much less sensitive to impurities. The introduction of carbon enhances mostly the intraband scattering in the $\sigma$-band. 
In contrast to the previously observed anomalous behavior of pure MgB$_2$, the Wiedemann-Franz law is valid for Mg(B$_{0.94}$C$_{0.06}$)$_2$ at low temperatures.  
\end{abstract}
\maketitle

The superconducting state of MgB$_2$ is characterized by at least two gaps in the electronic excitation spectrum, both of significantly different width (for a review, see Ref.~\onlinecite{TwoGaps} and references therein). 
The larger energy gap, $\Delta_{\sigma}$, develops on quasi-two dimensional (2D) sheets of the Fermi surface, often denoted as $\sigma$-band. A distinctly smaller gap $\Delta_{\pi}$ is formed on 3D parts of the Fermi surface that are related with the so-called $\pi$-band.
Amazingly, lattice disorder, which is thought to lead to interband scattering and consequently to an equalization of gap amplitudes\cite{Liu01} has very little effect on the two-gap nature of superconductivity in MgB$_2$. 
For example, the residual resistivity $\rho_0$, which provides a measure of impurity scattering, varies by orders of magnitude in different MgB$_2$ samples but no significant concomitant variation of T$_c$ is observed.  
The difference in the symmetry of the orbitals forming the quoted bands in MgB$_2$ was suggested to be the reason for weak {\em interband} scattering.\cite{Mazin02_Sup} 
The {\em intraband} scattering rates in each band may lead to different mean free paths in each band but will not affect the multigap nature of the superconducting state. 
Current studies of MgB$_2$ aim at a controlled variation of the influence of interband and intraband scattering processes on physical properties of MgB$_2$. 
One way of achieving this is provided by selected small variations of the chemical composition, leading to selective variations in the quoted scattering processes. 
In particular, carbon substitution for boron is expected to enhance the intraband scattering rate mostly in the $\sigma$-band, which is formed by the boron $s p_x p_y$ orbitals, but it does not alter the interband scattering very much because carbon impurities do not generate considerable out-of-plane distortions.\cite{Erwin03}
Various experimental results support the persistence of two different gaps in Mg(B$_{1-x}$C$_x$)$_2$ up to  $x\approx 0.1$, \cite{Schmidt03,Papagelis03_EPL,Samuely03_Two,Holanova04,Gonnelli04cm} but conflicting results are reported for higher carbon concentrations.\cite{Holanova04,Gonnelli04cm} Carbon doping is also reported to cause a considerable increase of the upper critical field and some reduction of its anisotropy.\cite{Ribeiro03_Car,Pissas03cm,Puzniak04cm,Kazakov04cm} 

In previous work it was demonstrated that important aspects of multigap superconductivity in MgB$_2$ can experimentally be addressed by measurements of the thermal conductivity $\kappa$ as a function of temperature $T$ and external magnetic field $H$.\cite{Muranaka01,Bauer01,Putti01,Schneider01,Sologubenko02_Hc2,Sologubenko02_KH,Putti03} 
Particularly useful are $\kappa(H)$ data in the mixed state,\cite{Sologubenko02_KH}  because they 
allow for distinguishing between the  contributions of the different bands to the electronic heat transport and the scattering of phonons by electrons. 
Accompanying theoretical considerations\cite{Tewordt02,Kusunose02,Tewordt03_Two,Tewordt03_Spe} support
these interpretations.

In this paper, we present a set of $\kappa(T,H)$ data obtained for single crystals of Mg(B$_{1-x}$C$_x$)$_{2}$ ($x= 0.03, 0.06$) in the basal plane of the hexagonal crystal structure
and  we compare them with our previously published, analogous data for pure MgB$_2$. 
The crystals were grown with a high-pressure technique in a cubic anvil press as described in Ref.~\onlinecite{Kazakov04cm}. Two approximately bar-shaped single crystals with dimensions of
$0.8\times 0.27\times 0.10$ mm$^{3}$ for Mg(B$_{0.97}$C$_{0.03}$)$_2$
and 
$0.9\times 0.26\times 0.13$ mm$^{3}$ for Mg(B$_{0.94}$C$_{0.06}$)$_2$ were selected for the thermal conductivity measurements. 
For both samples, the shortest extension in size is along the $c$-axis.
The high structural quality and the homogeneity of the carbon distribution
were confirmed by single-crystal X-ray diffraction and EDX analyses 
on similar crystals from the
same batches. A standard uniaxial heat-flow method was used for  measuring $\kappa(T,H)$,
with the same experimental arrangements as reported in our previous works on pure MgB$_2$.\cite{Sologubenko02_Hc2,Sologubenko02_KH} The Mg(B$_{0.94}$C$_{0.06}$)$_2$ sample was also used for 4-contact $dc$ measurements of the electrical resistivity.

\begin{figure}[t]  
 \begin{center}
  \leavevmode
  \epsfxsize=0.8\columnwidth \epsfbox {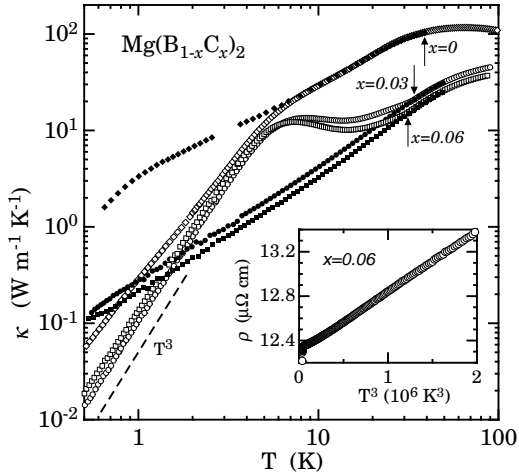}
   \caption{
Thermal conductivity $vs$ temperature in the $ab$-plane of single-crystalline Mg(B$_{1-x}$C$_x$)$_2$ ($x = 0, 0.03, 0.06$) in zero magnetic field (open symbols) and $H \parallel c = 50 {\rm ~kOe}$ (solid symbols). The arrows indicate the zero-field critical temperature $T_{c}$. 
The inset emphasizes the $T^3$ variation of the in-plane electrical resistivity $\rho(T)$ of  Mg(B$_{0.94}$C$_{0.06}$)$_2$.
  }
\label{KT}
\end{center}
\end{figure}

The temperature dependences of thermal conductivities of carbon-doped MgB$_2$ in zero magnetic field and $H=50 {\rm ~kOe}$ are presented in Fig.~\ref{KT}. For comparison, our earlier data\cite{Sologubenko02_KH} for pure MgB$_2$ are also shown.
The substitution of carbon for boron results in a considerable reduction of the zero-field thermal conductivity, mainly in the vicinity and above $T_c$.
For the C-doped samples we note a distinct change in $\partial\kappa/\partial T$
at $T_c$, in contrast to the previously noted, striking absence of any such feature at $T_c$ of undoped MgB$_2$.\cite{Muranaka01,Bauer01,Putti01,Schneider01,Sologubenko02_KH}     
%<<<<<<<<<<<<<<<<<<<<<<<< FIGURE 1 >>>>>>>>>>>>>>>>>>>>>>>>>
%<<<<<<<<<<<<<<<<<<<<<<<< FIGURE 1 >>>>>>>>>>>>>>>>>>>>>>>>>
%<<<<<<<<<<<<<<<<<<<<<<<< figure 1 >>>>>>>>>>>>>>>>>>>>>>>>>

\begin{figure}[t]
 \begin{center}
  \leavevmode
  \epsfxsize=1\columnwidth \epsfbox {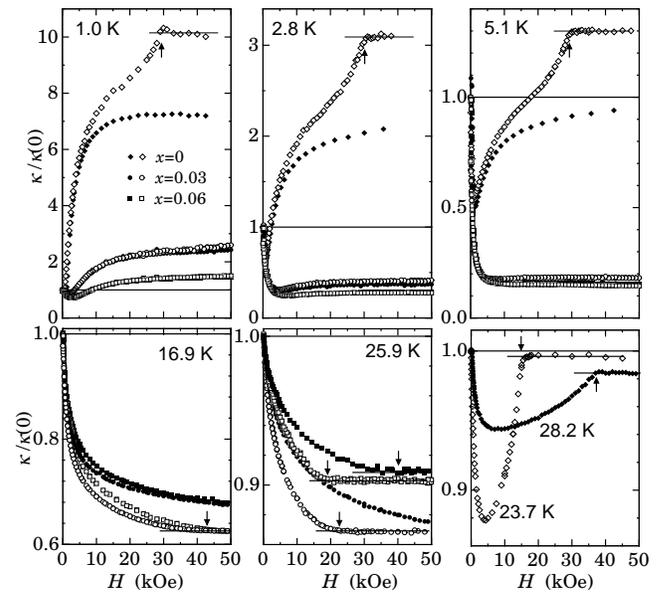}
   \caption{
  Thermal conductivity in the basal plane of Mg(B$_{1-x}$C$_x$)$_2$ ($x$=0, 0.03 and 0.06) vs $H$ at several fixed temperatures. The arrows indicate the upper critical field $H_{c2}$. The closed and open  symbols correspond to the field
  directions perpendicular and parallel to the $c-$axis, respectively.
  }
\label{KH}
\end{center}
\end{figure}

Selected $\kappa(H)$ data at several fixed temperatures for $x$=0, 0.03, and 0.06 are shown in Fig.~\ref{KH}.
At 0.6~K, $\kappa(H)$ is reversible above at least 1~kOe and at higher temperatures, the irreversibility field decreases. In what follows, we only discuss  the $\kappa(H)$ curves in the reversible region above $H_{c1}$.   
Increasing $H$  above $H_{c1}$ leads, at all temperatures, to an initial decrease of $\kappa$.  
With further increasing $H$, $\kappa$ increases and, when $H$ exceeds the upper critical field $H_{c2}$, the thermal conductivity is practically independent of field.
At $T \ll T_c$,  $\kappa(H)$  for pure MgB$_2$ depends very little on the orientation of the magnetic field below about 5-10~kOe, but in higher fields, the $\kappa$ values for $H \parallel c$ exceed those for $H \perp c$ significantly (see the upper row in Fig.~\ref{KH}). For the carbon-doped samples, $\kappa(H)$ depends only weakly on the field orientation in all fields. Another striking observation across the entire covered temperature range is the much weaker field-induced enhancement of $\kappa$ for the carbon-doped samples than for pure MgB$_2$. 
At temperatures above about 5~K and for $x=0.03$ and  $x=0.06$, no field-induced enhancement of $\kappa$ is observed at all, whereas for $x=0$, it persists at all temperatures up to $T_c$ (see the lower row in Fig.~\ref{KH}). 
%<<<<<<<<<<<<<<<<<<<<<<<< FIGURE 2 >>>>>>>>>>>>>>>>>>>>>>>>>
%<<<<<<<<<<<<<<<<<<<<<<<< FIGURE 2 >>>>>>>>>>>>>>>>>>>>>>>>>
%<<<<<<<<<<<<<<<<<<<<<<<< figure 2 >>>>>>>>>>>>>>>>>>>>>>>>>

\begin{figure}[t]
 \begin{center}
  \leavevmode
  \epsfxsize=0.8\columnwidth \epsfbox {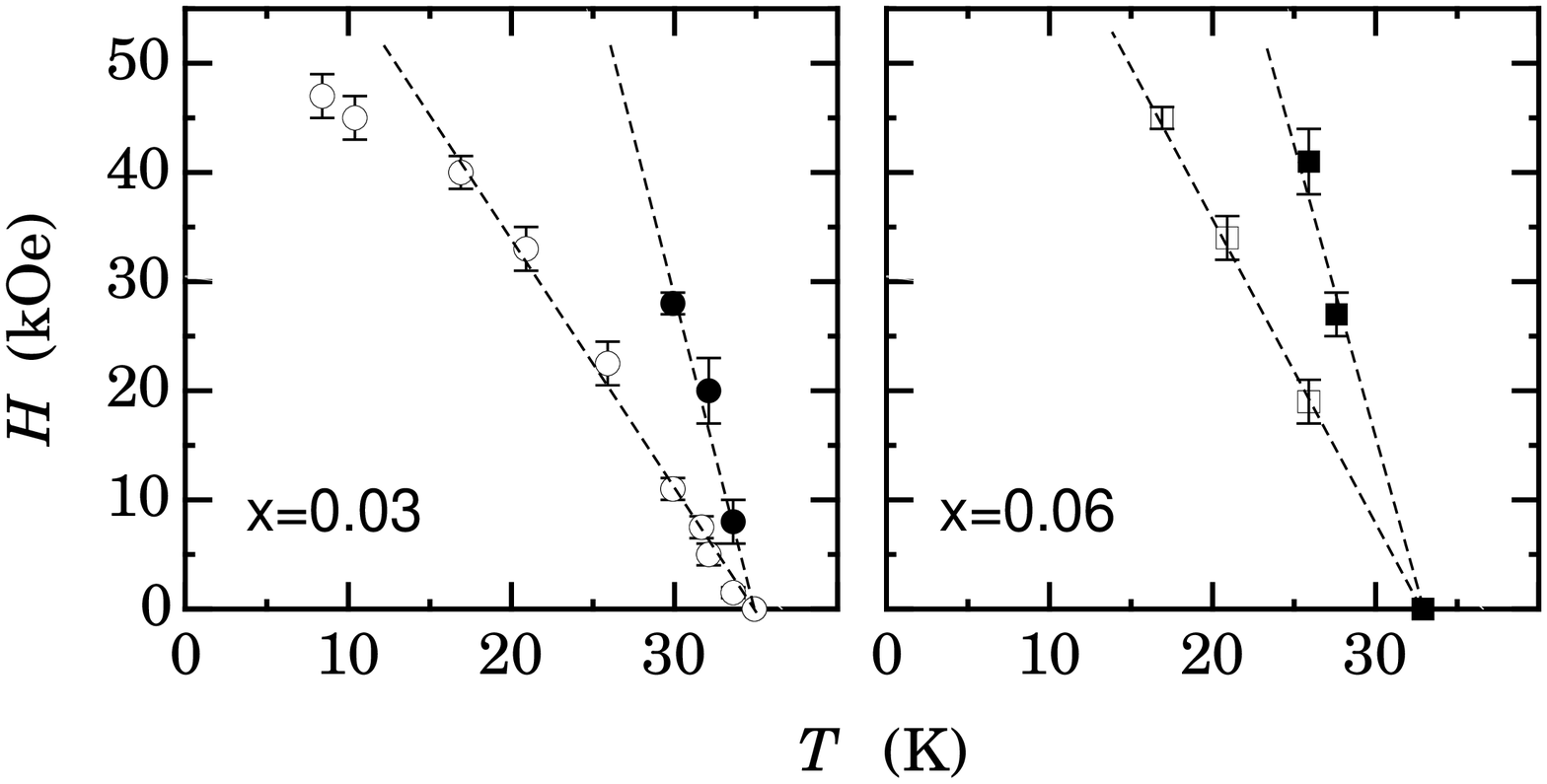}
   \caption{
  The upper critical fields $H_{c2}(T)$ as evaluated  from  thermal conductivity data (open symbols for $H \parallel c$ and solid symbols for $H \perp c$). 
The dashed lines are to guide the eye.
  }
\label{Hc2}
\end{center}
\end{figure}

The $\kappa(H)$ curves allow to establish the values of the bulk upper critical field $H_{c2}$ via the onset of the field independence of $\kappa(H)$ (see arrows in Fig.~\ref{KH}). 
The resulting $H_{c2}(T)$ data are shown in Fig.~\ref{Hc2}. They are in  agreement with results of torque measurements on similar samples.\cite{Puzniak04cm} 
%<<<<<<<<<<<<<<<<<<<<<<<< FIGURE 3 >>>>>>>>>>>>>>>>>>>>>>>>>
%<<<<<<<<<<<<<<<<<<<<<<<< FIGURE 3 >>>>>>>>>>>>>>>>>>>>>>>>>
%<<<<<<<<<<<<<<<<<<<<<<<< figure 3 >>>>>>>>>>>>>>>>>>>>>>>>>
 
The  zero-field electrical resistivity for the $x=0.06$ sample was measured between 4.2 and 130~K. 
The transition at $T_c$=32.9~K is, with a width of about 0.1~K, rather narrow.  
In the entire covered temperature range above $T_c$, $\rho(T)$ is well described by $\rho = \rho_0 + a T^3$ (see the inset in Fig.~\ref{KT}) 
where two scattering processes, associated with quasiparticle scattering by defects and phonons, respectively, are considered. The constant parameters are $\rho_0=12.3 {\rm ~\mu\Omega cm}$ and 
$a = 5.4 \times 10^{-7} {\rm ~\mu\Omega cm K^{-3}}$. In comparison, for pure MgB$_2$, $\rho_0=2.0 {\rm ~\mu\Omega cm}$ and 
$a = 6.7 \times 10^{-7} {\rm ~\mu\Omega cm K^{-3}}$ (Ref.~\onlinecite{Sologubenko02_KH}), which demonstrates that carbon doping strongly enhances the scattering by defects, while the electron-phonon scattering is only weakly altered.

Because we assume that the transport is provided by itinerant electronic ($\kappa_e$) and lattice excitations ($\kappa_{\rm ph}$), such that $\kappa = \kappa_e + \kappa_{\rm ph}$, any analysis of the changes of $\kappa(T,H)$  caused by the substitution of carbon for boron should consider corresponding changes in both these channels.
In MgB$_2$, the electronic thermal conduction consists of contributions from both the $\sigma$- and the $\pi$-band, hence $\kappa_e = \kappa_{e,\sigma} + \kappa_{e,\pi}$.\cite{Tewordt02,Kusunose02,Tewordt03_Two,Tewordt03_Spe} 
When the temperature is reduced to below $T_c$ in zero magnetic field, the continuous reduction of the number of unpaired quasiparticles leads to a decrease of $\kappa_e$ but, because of the reduction of the scattering of phonons by electrons (holes), also to an increase of $\kappa_{\rm ph}$.
Altogether, the features of $\kappa(T)$ depend on the relative magnitudes of $\kappa_e$ and $\kappa_{\rm
ph}$ and on the efficiency of phonon scattering by electrons. Since for MgB$_2$ the energy gap in the $\pi$-band at a given normalized temperature $T/T_c$ is considerably smaller than expected from the BCS-theory, the $\pi$-band contributions to both the heat transport and the phonon scattering extend to lower temperatures than in conventional superconductors. 
Nevertheless, both $\kappa_{e,\sigma}$ and $\kappa_{e,\pi}$ are negligible at very low temperatures and $H=0$. 
Below 2-3~K, the observed $\kappa(T,0)$ varies approximately as  $T^3$ (see dashed line in Fig.~\ref{KT}), which is expected for purely phononic conductivity and a dominant scattering of phonons at the sample boundaries. 
The observation that the $\kappa(T,0)$ curves tend to merge below $T_c$ suggests that carbon doping does not influence the thermal conduction via phonons very much, and that the difference in $\kappa(T,0)$ observed at higher temperatures results from $\kappa_e$ being considerably reduced by the introduction of carbon for boron.
The question of which band is more strongly influenced by doping, can be answered by analyzing the $\kappa(H)$ data in the mixed state.

The quasiparticles associated with the vortices provide
additional scattering of phonons and also enhance $\kappa_{e}$. The competition of these two mechanisms results in typical $\kappa(H)$ curves for type-II superconductors, with an initial drop of $\kappa$ when $H$ is raised to above $H_{c1}$ and subsequently to a considerable increase when $H$ approaches $H_{c2}$. 
For MgB$_2$ this rather gradual increase of $\kappa_e(H)$ is replaced by an extremely rapid increase of $\kappa_e$ at fields $H \ll H_{c2}$, followed by a saturation region and yet another increase in the vicinity of $H_{c2}$.\cite{Sologubenko02_KH} 
The low-field increase does not scale with $H/H_{c2}$ as is observed for conventional superconductors and is almost independent of field orientation, despite the strong anisotropy of $H_{c2}$. 
This unique behavior is explained in terms of a two-band scenario, where the smaller energy gap $\Delta_{\pi}$ is closed at a relatively low and weakly-anisotropic field $H^*_{\pi}<H_{c2}$, of the order of 10~kOe for pure MgB$_2$.\cite{Bouquet02,Samuely03_Poi,Lyard04}  
Therefore, $\kappa_{e,\pi}$ increases rapidly when the field approaches $H^*_{\pi}$ and, still in the mixed state, adopts its normal-state value above $H^*_{\pi}$. The $\sigma$-band contribution $\kappa_{e,\sigma}$ only grows when $H$ approaches $H_{c2}$, the enhancement being strongly anisotropic, reflecting the anisotropy of $H_{c2}$. 

The striking difference between carbon-free and carbon-doped samples is that, for the latter, no significant increase in $\kappa(H)$ is observed in the vicinity of $H_{c2}$. It implies that $\kappa_{e,\sigma}$ is strongly reduced  by carbon substitution  from about 50\% of the total normal-state $\kappa_e$ for $x=0$ (Ref.~\onlinecite{Sologubenko02_KH}) to a level close to our experimental resolution for $x=0.03$. Hence, the quasiparticle contribution to the heat transport is dominated by the $\pi$-band, i.e., $\kappa_e \approx \kappa_{e,\pi}$.  
Carbon doping also reduces $\kappa_{e,\pi}$ but less severely than $\kappa_{e,\sigma}$. 
At 1.0~K and in a field $H\perp c$=45~kOe,  the thermal conductivity
for the $x=0.06$ sample is a factor of 11 smaller than for pure MgB$_2$. Since at such low temperatures and high fields $\kappa_{\rm ph}$ is rather small, this reduction is obviously due to the carbon-doping induced reduction of $\kappa_{e,\pi}$.  
The quasiparticle thermal conductivity is given by $\kappa_{e} = C_{e} v_{F} 
\ell/3$, where $C_e$  is the electronic specific 
heat, $v_F$ the Fermi velocity and $\ell$  the mean free path. 
If we assume that by carbon doping at the 6\% level neither the electronic density of 
states at the Fermi level $N_{\pi}(E_F)\propto C_{e,\pi}$
nor the Fermi velocity $v_{F,\pi}$ in the $\pi$-band are much altered, then the reduction of $\kappa_{e,\pi}(x)$ simply reflects the reduction of the corresponding mean free path $\ell_{\pi}$, due to enhanced scattering by carbon.  The reduction of $\kappa_{e,\sigma}$ is at least an order of magnitude stronger, implying a corresponding reduction of the mean free path $\ell_{\sigma}$. 
With the same assumptions, the reduction of the mean free path may also be evaluated from the values of the residual electrical resistivity $\rho_0$. Since both bands contribute to electrical conductivity, the factor  $\rho_0(x=0)/\rho_0(x=0.06) \approx 6$ represents the reduction of the mean free path, averaged for both bands.
The value of this factor is of the same order of magnitude as calculated from $\kappa_{e,\pi}$.
This observation corroborates theoretical work of Mazin {\em et al.}\cite{Mazin02_Sup} which suggests that the residual resistivity of MgB$_2$ is mostly due to the intraband scattering of holes in the $\pi$-band.    

In previous work\cite{Sologubenko02_KH} we reported a significant violation of the Wiedemann-Franz law (WFL) in the field-induced normal state of MgB$_2$ at low temperatures, below about 6~K. 
The WFL relates electrical and thermal transport by  
\begin{equation}\label{eWFL}
\kappa_{e}(T) = L_{0} T / \rho(T),
\end{equation}
where $L_{0}=2.45\times 10^{-8}$ W~$\Omega$ K$^{-2}$. This relation normally holds for common metals if the relaxation rates for electrical and thermal currents 
are equal. Elastic scattering of electrons by defects usually guarantees this equality and hence, in the region where this scattering dominates, in particular where $\rho(T) \approx \rho_{0}$, the validity of the WFL is expected. 
The Lorenz number $L(T) = \kappa(T) \rho(T) T^{-1}$ where $\kappa(T)$ is the total measured conductivity, is equal to $L_0$ only if the phonon thermal conductivity is negligible. 
In the low-temperature normal state of MgB$_2$, induced by fields $H \parallel c > 30 {\rm ~kOe}$,  the ratio $L(T)/L_0$ considerably exceeds unity and exhibits a peak at about 1~K, where $L/L_0 \approx 3$.\cite{Sologubenko02_KH} Our estimates for the upper limit of $\kappa_{\rm ph}$  at these temperatures revealed that the phonon contribution to $\kappa(T)$ cannot be responsible for the enhanced Lorenz number, but the cause for the violation of the WFL is not yet firmly established. 

\begin{figure}[t]
 \begin{center}
  \leavevmode
  \epsfxsize=0.8\columnwidth \epsfbox {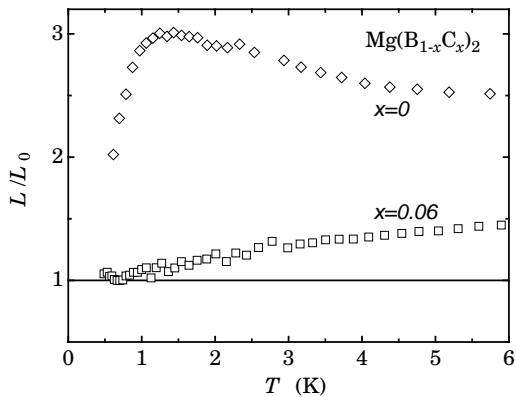}
   \caption{Normalized Lorenz number $L(T)/L_0$ for Mg(B$_{1-x}$C$_x$)$_2$ ($x=0$ and 0.06).}
\label{LL0}
\end{center}
\end{figure}

For the 6\% carbon-doped MgB$_2$, we cannot exceed the upper critical field at low temperatures with our experimental facilities. A direct evaluation of $L(T)$  in the normal state is thus not possible. However, since in our low-temperature data the quasiparticle contribution $\kappa_e(H)$ saturates to the normal-state values already in the mixed state above $H^*_{\pi}$, $L(T)$ data at fields $H \gg H^*_{\pi}$ can be used. In Fig.~\ref{LL0}, the low-temperature values of $L(T)/L_0$ at $H=50 {\rm ~kOe}$ are presented for the $x=0.06$ sample and, for comparison, for the undoped MgB$_2$.\cite{Sologubenko02_KH} 
In both cases, $L(T)$ was calculated as discussed in Ref.~\onlinecite{Sologubenko02_KH}.
In contrast to pure MgB$_2$, $L/L_0 \approx 1$ for Mg(B$_{0.94}$C$_{0.06}$)$_2$   at temperatures $T \lesssim 1 {\rm ~K} $. The small deviation at elevated temperatures can naturally be attributed to $\kappa_{\rm ph}$ which, with decreasing temperature, becomes much smaller that $\kappa_e$. Thus, at least at low temperatures, no violation of the WFL is observed in the carbon-doped MgB$_2$. It is important to note that in this case, the  heat transport is predominantly carried by the quasiparticles of the quasi-3D $\pi$-band. Since in pure MgB$_2$, where the violation of the WFL is pronounced, both the $\sigma$- and the $\pi$-band  contribute almost equally to the normal-state heat transport, it is natural to conclude that the previously observed anomaly is  predominantly tied to the transport in the quasi-2D $\sigma$-band. 
Our results cannot really identify the origin of the violation of the WFL, but they do single out the involved electronic states that are responsible for the effect.
%<<<<<<<<<<<<<<<<<<<<<<<< FIGURE 4 >>>>>>>>>>>>>>>>>>>>>>>>>
%<<<<<<<<<<<<<<<<<<<<<<<< FIGURE 4 >>>>>>>>>>>>>>>>>>>>>>>>>
%<<<<<<<<<<<<<<<<<<<<<<<< FIGURE 4 >>>>>>>>>>>>>>>>>>>>>>>>> 

In conclusion, the substitution of carbon for boron  leads to a considerable reduction of the electronic thermal conductivity of MgB$_2$, whereby the transport in the $\sigma$-band is much more strongly affected than that in the $\pi$-band. The Wiedemann-Franz law, considerably violated for pure MgB$_2$ at low temperatures, seems to be valid for Mg(B$_{0.94}$C$_{0.06}$)$_2$.

\acknowledgments
This work was financially supported in part by
the Schweizerische Nationalfonds zur F\"{o}rderung der Wissenschaftlichen
Forschung.

\end{document}